\documentclass[%
preprint,
superscriptaddress,
 amsmath,amssymb,
 aip,
]{revtex4-1}

\usepackage{graphicx}
\usepackage{dcolumn}
\usepackage{bm}
\usepackage{braket}
\usepackage{siunitx}
\usepackage{hyperref}
\usepackage[mathlines]{lineno}

\begin{document}


\title{Tuning the carrier tunneling in a single quantum dot with a magnetic field in Faraday geometry}

\author{Kai Peng}
\author{Shiyao Wu}
\author{Xin Xie}
\author{Jingnan Yang}
\author{Chenjiang Qian}
\author{Feilong Song}
\author{Sibai Sun}
\author{Jianchen Dang}
\author{Yang Yu}
\author{Shan Xiao}
\affiliation{Beijing National Laboratory for Condensed Matter Physics, Institute of Physics, Chinese Academy of Sciences, Beijing 100190, China}
\affiliation{CAS Center for Excellence in Topological Quantum Computation and School of Physical Sciences, University of Chinese Academy of Sciences, Beijing 100049, China}
\author{Xiulai Xu}%
\email{xlxu@iphy.ac.cn}
\affiliation{Beijing National Laboratory for Condensed Matter Physics, Institute of Physics, Chinese Academy of Sciences, Beijing 100190, China}
\affiliation{CAS Center for Excellence in Topological Quantum Computation and School of Physical Sciences, University of Chinese Academy of Sciences, Beijing 100049, China}
\affiliation{Songshan Lake Materials Laboratory, Dongguan, Guangdong 523808, China}

\date{\today}

\begin{abstract}
We report on an increase in the carrier tunneling time in a single quantum dot (QD) with a magnetic field in Faraday geometry using photocurrent spectroscopy.
A nearly 60\% increase in hole tunneling time is observed with an applied magnetic field equal to 9 T.
For a truncated pyramid QD, hole tunnels out faster at the lateral edge of the QD due to the reduced barrier height.
The magnetic field in Faraday geometry shrinks the hole wave function at the center of QD plane, which weakens the tunneling at lateral edge and
increases the average tunneling time.
This mechanism also works for electron but the effect is smaller. The electron wave function is more localized at the center of the QD due to the uniform confining potential, therefore the relatively weak shrinkage caused by the magnetic field does not reduce the tunneling rate significantly.
\end{abstract}

\maketitle

Single spins of electron or hole trapped in semiconductor quantum dots (QDs) have been proved to be a promising qubit candidate for optical coherent control \cite{loss1998quantum,xu2007fast,Berezovsky349,press2008,brunner2009coherent,warburton2013single,ethier2017improving}. In particular, the heavy hole with a p-type wave function distribution has a long coherence time due to the small hyperfine interaction with nuclear spins \cite{bulaev2005spin,gerardot2008optical,brunner2009coherent,warburton2013single}.
Controlling the wave functions of carriers in QDs plays a key role in the implementation of solid-state quantum information processing and spintronics. For example, through manipulating the electronic wave functions with an applied lateral electric field, the exciton coherence time can be enhanced \cite{moody2016electronic}. In addition,
it is also an effective way to control the wave functions in QDs with magnetic fields. In the classical picture, the magnetic field applies a Lorentz force to the carrier perpendicular to the direction of motion of the carrier. The cyclotron motion can shrink the extent of the carrier wave function in the plane perpendicular to the magnetic field, which has been observed by mapping the wave function of electron through magnetotunneling spectroscopy in atom-like systems \cite{patan2010manipulating,lei2010artificial}. Controlling the wave function by magnetic fields has been used to tune the charge state \cite{moskalenko2008effective,tang2014charge} and the dipole moment \cite{cao2015longitudinal} in single QDs.

The tunneling of the carriers in QDs contributes to the current signal in the presence of high electric field, which can be measured by the photocurrent spectroscopy. Due to the high detection efficiency \cite{ramsay2010review}, the photocurrent spectroscopy has been utilized as an effective way to detect the charge- and spin-based qubit in single QDs \cite{zrenner2002coherent,Ramsay2008Fast,godden2012coherent}.
However, the tunneling of the carriers reduces the coherence time, which limits the practical application of the QD-based qubit. Therefore, prolonging the tunneling time of carriers in QDs can improve the performance of spin-based qubit in the coherent control with photocurrent spectroscopy.

In this paper, we investigate the effect of magnetic field in Faraday geometry on the tunneling of carriers in single QDs, and propose a mechanism with considering the wave function distribution and the structure of QDs. Here, the hole tunneling time is found increasing with an increase in magnetic field through the reduced saturation photocurrents. For electron, a smaller increase in the tunneling time is observed through the broadening of the homogeneous linewidth of photocurrent spectra. These effects are attributed to the shrinkage of the wave functions of carriers in the truncated pyramid QD under the magnetic field in Faraday geometry.
The electron or hole tunnels fast at the lateral edge of the QD due to the reduced barrier height. While the tunneling in this part is weakened as the wave function shrinks, resulting in an increased tunneling time. For the hole, the wave function is more localized at the bottom and spreads to the lateral edge of the QD. As a result, the tunneling rate reduction is more obvious than that of electron.

\begin{figure}
\includegraphics[scale=0.8]{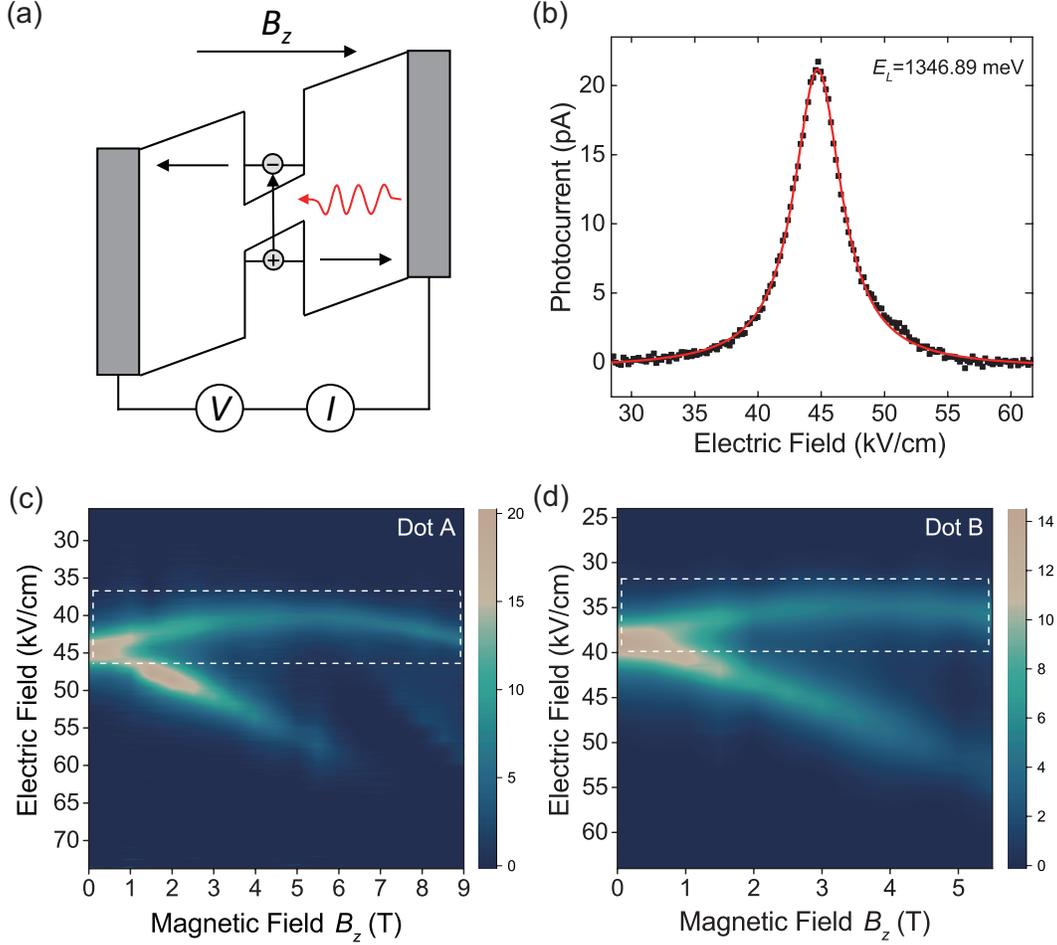}
\caption{\label{fig:1}(a) The photocurrent diagram of neutral exciton. (b) The photocurrent spectrum of dot A with a pumping laser as $E_L$=1346.89 meV. The red solid line is the Lorentzian fit to the experimental data. (c) and (d) The magneto-photocurrent spectra of single Dot A and Dot B in Faraday geometry ($B_z$). The linearly polarized laser is used to excite two branches of neutral exciton resonantly. The colorbar has the unit as picoampere. At 6 T, another QD's photocurrent peak appears in (c).
}
\end{figure}

The \emph{n-i-}Schottky device used in this work is designed for photocurrent measurement of single QDs. A single layer of InAs QDs with a low density ($\sim$10$^{9}$ cm$^{-2}$) is grown by molecular beam epitaxy along the [001] crystallographic direction and embedded in a 250-nm intrinsic GaAs layer. A two-dimensional electron gas is formed by a Si $\delta$-doped GaAs layer with a doping density $N_d$ = 5$\times$10$^{12}$ cm$^{-2}$. This layer is 50 nm below the QDs and is in contact with (Au, Ge)Ni alloy Ohmic contact. At the top of the structure, a 10-nm semitransparent Ti is evaporated on the GaAs surface to form a Schottky contact, followed by an Al mask with apertures of about 1-3 $\mu$m to isolate single QDs.
Finally, Cr/Au bond pads are formed on the Ohmic and Schottky contacts for connection to the external circuit.
The device is placed in a helium gas exchange cryostat at 4.2 K equipped with superconducting magnet capable of applying a magnetic field up to 9 T in Faraday geometry.
A tunable narrow-bandwidth ($\sim$1 MHz) external cavity diode laser in Littrow configuration is furnished in a confocal microscopy system to achieve resonant excitation of single QDs.
More details about the device and measurement can be found in Ref. \cite{peng2017probing,peng2019giant}.

The photocurrent measurement scheme is shown in Fig.~\ref{fig:1}(a). The resonant excitation of the neutral exciton generates an electron-hole pair in the QD s-shell. With an applied reverse bias voltage (high electric field), the electron and hole will tunnel out of the QD before recombination and contribute to measurable current signals added to the background.
In the experiment, the transition energy of neutral exciton is determined firstly through the bias-dependent photoluminescence spectroscopy. The photocurrent spectrum is measured by sweeping the neutral exciton energy through quantum-confined Stark effect to resonate with a wavelength-fixed laser.
A typical photocurrent spectrum (Dot A) is shown in Fig.~\ref{fig:1}(b).
The Lorentzian fit of the photocurrent spectrum gives the photocurrent amplitude, linewidth, and the corresponding central electric field.
Here, the electron will tunnel out first within several picoseconds due to the small effective mass.
As the main decoherence mechanism, the electron tunneling determines the homogeneous linewidth of the photocurrent spectrum \cite{stufler2004power}.
Meanwhile the measured linewidth also includes the contribution of power broadening, which is caused by the fact that the photocurrent amplitude is more easily saturated at exact resonance than off-resonance.
In contrast, the hole has a heavier effective mass and tunnels much slower than the electron, with a typically tunneling time of several nanoseconds.
Until the hole tunnels out, the next electron-hole pair can not be excited due to the mismatch of the transition energy between the neutral and the positive charged excitons.
As a result, the photocurrent amplitude is restricted by the hole tunneling.

\begin{figure}
\includegraphics[scale=0.8]{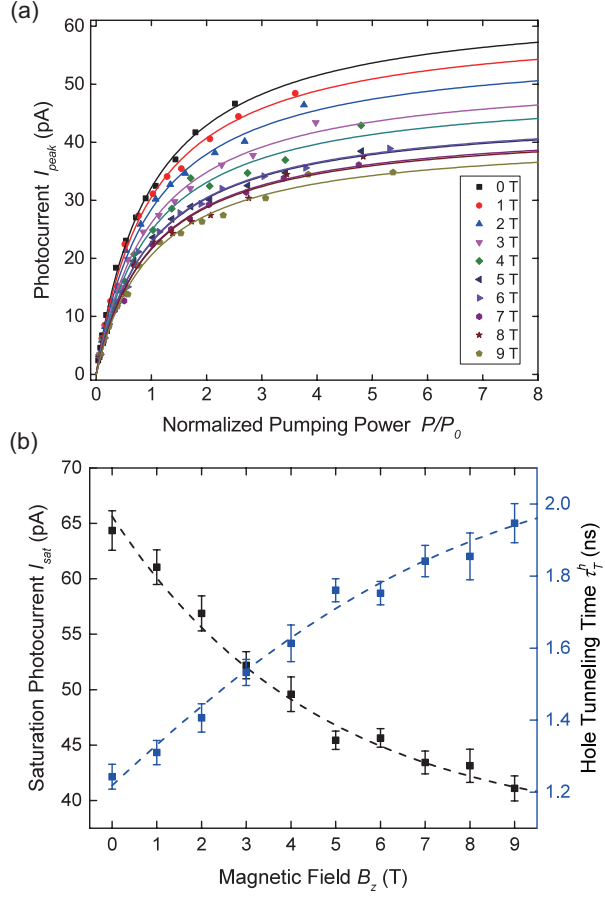}
\caption{\label{fig:2}(a) The pumping-power-dependent photocurrent amplitudes with an applied magnetic field from 0 to 9 T in Faraday geometry. The solid lines are the fits according to Eq.~(\ref{satur}). (b) The extracted saturation photocurrent and hole tunneling time.
}
\end{figure}

With an applied magnetic field in Faraday geometry, the photocurrent spectrum of neutral exciton shows the Zeeman splitting and the diamagnetic effect (moving to high electric field), as shown in the photocurrent spectra of Dot A and Dot B from two different Schottky devices in Fig.~\ref{fig:1}(c) and ~\ref{fig:1}(d), respectively. In order to observe the two splitting branches with orthogonal circular polarizations, the linearly polarized laser is chosen for the excitation. It is obvious that the tunneling rates of carriers depend on the electric field \cite{beham2001nonlinear,mar2011voltage,mar2011electrically}. As a result, the linewidth of the photocurrent spectrum gets broader and the amplitude gets larger as the peaks moving to high electric field.
Meanwhile, the magnetic field also changes the tunneling rates of carriers and affects the amplitude and the linewidth of the photocurrent spectrum. For the upper branches of Dot A and Dot B, as shown in the dashed rectangles in Fig.~\ref{fig:1}(c) and ~\ref{fig:1}(d), the photocurrent amplitudes decrease with the increase of the magnetic field under relatively stable electric fields. Here, we focus on the adjustments to the tunneling rates of carriers caused by magnetic fields, so the upper branch of Dot B is chosen to avoid the interferences from the electric field and other QDs.
The pumping laser wavelength is tuned for the following measurements to make sure the central electric field of the photocurrent spectrum is located at 41.3$\pm$0.2 kV/cm.

As mentioned above, the hole tunneling rate limits the photocurrent amplitude. With an increase in the pumping power, the photocurrent amplitude saturates. Through the saturation photocurrent amplitude, the hole tunneling time can be calculated quantitatively. The pumping-power-dependent behavior can be described by the following theoretical model based on a two-level system \cite{beham2001nonlinear}:
\begin{equation}
I_{peak}=I_{sat}\frac{P}{P+P_0}=\frac{e}{2\tau^h_T}\frac{P}{P+P_0},\label{satur}
\end{equation}
where \emph{P} is the pumping power on the QD, \emph{e} is the elementary charge, and $\tau^h_T$ is the hole tunneling time. $P_0$ is the power required for saturation, which is proportional to the hole tunneling time. Figure.~\ref{fig:2}(a) shows the photocurrent amplitudes as a function of the pump power, under an  applied magnetic field changing from 0 to 9 T. To avoid the influence of the other branch, circularly polarized laser is chosen to only excite the upper branch. The saturation photocurrent decreases as the magnetic field increases. The hole tunnelling time is derived from Eq.~(\ref{satur}), as shown in Fig.~\ref{fig:2}(b). For this QD, the hole tunneling time increases monotonically.
For example, the tunneling time increases by nearly 60\%, from 1.2 ns at zero magnetic field to about 2 ns at 9 T.

\begin{figure}
\includegraphics[scale=0.8]{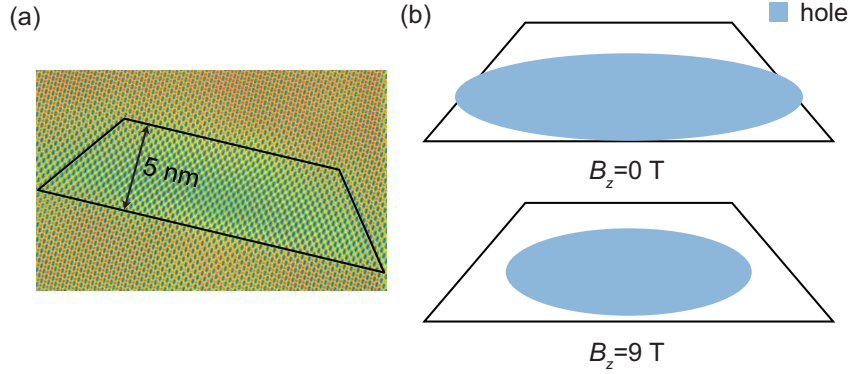}
\caption{\label{fig:3}(a) A high-resolution cross section image of a single QD along [110] crystallographic direction by transmission electron microscope. (b) The schematic diagrams of hole wave functions at ground state for an applied magnetic field of 0 T and 9 T in Faraday geometry.
}
\end{figure}

When the magnetic field is applied perpendicular to the tunneling direction, i.e. in Voigt geometry, the Lorentz force deflects the momentum of carriers along the tunneling direction and decreases the tunneling rates of carriers \cite{patane2002probing,godden2012fast}. Based on this mechanism, the magnetotunneling spectroscopy has been used to map the electron wave function in QDs in momentum space by current and capacitance measurements \cite{vdovin2000imaging,patane2002probing,wibbelhoff2005magneto}.
While for the magnetic field in Faraday geometry, paralleling to the tunneling direction, the induced cyclotron motion of the carrier is in the QD plane, which does not affect the vertical momentum directly. However, considering the QD structure, this phenomenon can be explained from the perspective of wave function distribution. The QDs in this sample have a truncated pyramid structure, as shown in the high-resolution cross section image of a single QD by transmission electron microscopy in Fig.~\ref{fig:3}(a) \cite{cao2015longitudinal}.
For the heavy hole in pure InAs QDs, the confining potential increases linearly from the bottom to the apex based on the eight-band \textbf{k$\cdot$p} calculation \cite{sheng2001electron,sheng2003absence}. As a result, hole wave function occupies the bottom of the QDs and spreads to the lateral edges, as shown schematically in the top panel of Fig.~\ref{fig:3}(b).
The hole tunneling is the integral effect of the distribution of the whole wave function. At the lateral edge of the QD, the hole tunnels fast due to the relatively smaller tunneling barrier height.

When the magnetic field is applied in Faraday geometry, the hole wave function shrinks to the middle because of the cyclotron motion, as shown schematically in the bottom panel of Fig.~\ref{fig:3}(b). As a result, the tunneling at the lateral edge is weakened and the average tunneling time increases.
The magnetic-field-induced shrinkage of the wave function can be described by the magnetic length \cite{davies1998physics,tsai2008diamagnetic}:
$l_B=\sqrt{\hbar/eB}$,
which characterizes the confinement of the magnetic field to the carrier. At $B_z$=1 T, the magnetic length of 26 nm is comparable to the lateral size of the QD.
In this case, the magnetic field does not provide more lateral confinements to the hole, which corresponds to a weak shrinkage of the hole wave function. When $B_z$=9 T, the corresponding magnetic length is 8.6 nm. Therefore the magnetic confinement dominates and induces the hole wave function to shrink to the center of the QD.
Due to relatively constant height of the QD in the middle, the effective barrier will not change significantly with the further shrinkage of the hole wave function. Therefore, the hole tunneling time saturates slightly as shown in Fig.~\ref{fig:2}(b).
It is worth noting that, the longitudinal wave function control of the hole by a magnetic field brings another effect that the center of the hole wave function moves towards to the apex of the QD. The resulting inverted permanent dipole moment of doubly negatively charged exciton has been observed by photoluminescence spectroscopy in our previous work \cite{cao2015longitudinal}. Even though the mechanism of the shrinkage of the wave function caused by the magnetic field in Faraday geometry is the same, here the tunneling process is considered.

\begin{figure}
\includegraphics[scale=0.8]{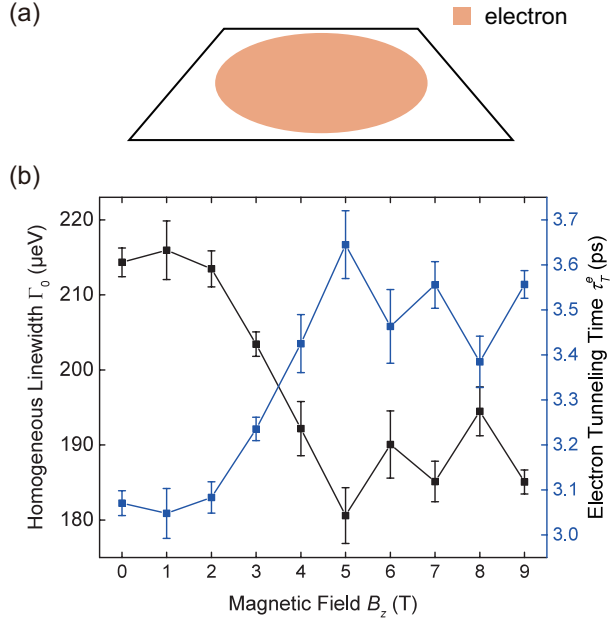}
\caption{\label{fig:4}(a) The schematic diagram of electron wave function at ground state. (b) The homogeneous linewidth of photocurrent spectra and extracted electron tunneling time with an applied magnetic field in Faraday geometry from 0 to 9 T.
}
\end{figure}

In contrast, the confining potential for electron is relatively uniform throughout the structure, resulting in a electron wave function distribution to be close to the center of the QD \cite{sheng2001electron,sheng2003absence}, as shown in Fig.~\ref{fig:4}(a). As a result, the neutral exciton has a dipole moment along the direction from the bottom toward the apex. The value \emph{p/e} is measured as 0.29 nm for Dot B through Stark effect \cite{peng2017probing}.
In order to analyze the electron tunneling process under the magnetic field, the linewidth of the photocurrent spectrum is used to determine the electron tunneling time, which is obtained through the relation between the transition energy of neutral exciton and the electric field according to the Stark effect \cite{peng2017probing}.
To derive the homogeneous linewidth, the following model is used to describe the power-dependent linewidth behavior \cite{allen1987optical,stufler2004power}:
\begin{equation}
\Gamma=\Gamma_0\sqrt{P/P_0+1}=\frac{\hbar}{\tau^e_T}\sqrt{P/P_0+1},\label{linewidth}
\end{equation}
where $\Gamma_0$ is the homogeneous linewidth and $\tau^e_T$ is the electron tunneling time. With the same $P_0$ fitted from the saturation photocurrent for each magnetic field, the fitted homogeneous linewidth and electron tunneling time are shown in Fig.~\ref{fig:4}(b). Even though with relatively large error, the electron tunneling time shows an increase with increasing magnetic field. This means that the shrinkage of the electron wave function also exists.
Since the electron wave function is already localized in the center in the absence of a magnetic field, the shrinkage caused by the magnetic field of 9 T is relatively weak. As a result, the electron tunneling time does not increase significantly.
In the experiment, the increase of electron tunneling time is only about 15.8\% from 0 to 9 T, which is much smaller than that of hole. The different modulations of the magnetic field on the tunneling time of electron and hole further confirm our model with considering the different wave function distributions.

The effect of the magnetic field in Faraday geometry on the tunneling of electron from the emitter to the discrete states has been reported in self-assembled QDs by capacitance spectroscopy \cite{lei2010artificial}, as well as in hydrogenic Si donors by tunneling spectroscopy \cite{patan2010manipulating}. The shrinkage of the electron wave function can be mapped with applied in-plane magnetic fields (Voigt geometry). While these measurements are averages for ensembles, and the p-shell electrons in QDs involve Coulomb interaction between electrons \cite{rontani2011artificial}.
Compared with the tunneling from the emitter to the atom-like solid-state systems in previous works \cite{patan2010manipulating,lei2010artificial}, here, we quantitatively investigated the tunneling of photogenerated carriers by focusing on the electron and hole in the s-shell in single QDs.
Even though the increased confinement caused by the magnetic field is common in low-dimensional systems, the truncated pyramid structure of the QD with different wave function distributions of electron and hole has to be considered to explain the decrease of the tunneling rate of carriers in single QDs.

In summary, we have demonstrated the modulations of the tunneling time of carriers in single QDs by the magnetic field in Faraday geometry using photocurrent spectroscopy. It is explained qualitatively by the combination of the wave function shrinkage and the truncated pyramid structure.
The magnetic field in Faraday geometry shrinks the wave functions of electron and hole, which weakens the tunneling at the lateral edge in the QD with a relatively smaller barrier height. Considering the different wave function distributions, the modulation of the tunneling time is more obvious for hole than electron.
Tuning the tunneling time of carriers in QDs through a longitudinal wave function modulation by the magnetic field provides a way to prolong the coherence time of hole-based qubit in photocurrent experiments. Furthermore, it is also helpful to understand the tunneling process in other low-dimensional structures from the wave function distribution perspective.
\\

This work was supported by the National Natural Science Foundation of China under Grants No. 61675228, No. 11721404, No. 51761145104 and No. 11874419; the Strategic Priority Research Program, the Instrument Developing Project and the Interdisciplinary Innovation Team of the Chinese Academy of Sciences under Grants No. XDB07030200, No. XDB28000000 and No.YJKYYQ20180036. We thank Weidong Sheng for helpful discussions.


\nocite{*}
\providecommand{\noopsort}[1]{}\providecommand{\singleletter}[1]{#1}%

\end{document}